\documentclass[10pt,journal,compsoc]{IEEEtran}

\ifCLASSOPTIONcompsoc
  \usepackage[nocompress]{cite}
\else
  \usepackage{cite}
\fi
\ifCLASSINFOpdf
\else
\fi
\usepackage{amsmath}
\begin{document}
%
\title{The relation between eigenvalue/eigenvecor and matrix game}
%
%
%
%

\author{Haolin Liu}

\IEEEtitleabstractindextext{%
\begin{abstract}
Matrix game, which is also known as two-person zero-sum game, is a famous model in game theory. There are some well-estabilished theory about it, such as von Neumann's minimax theorem. However, almost no literature have reported the relationship between eigenvalue/eigenvector and properties of matrix game. In this paper, we find such relation of some special martices and try to extend some conclusions to general matrix. 
\end{abstract}

}

\maketitle

\IEEEdisplaynontitleabstractindextext

%
\IEEEpeerreviewmaketitle

\ifCLASSOPTIONcompsoc
\IEEEraisesectionheading{\section{Introduction}\label{sec:introduction}}
\else
\section{Introduction}
\label{sec:introduction}
\fi

%
%
%
%
\IEEEPARstart{M}{atrix} games are two-person zero sum game with finite startegy sets. Despite its simple structure, matrix games often appear in real life, the well-known game Rock-paper-scissor is a matrix game. Having a good understanding of matrix game is useful when making choice, thus many scientists have done a lot of deep research in this field, but few of them have studied the relation between matrix game and eigenvalue/eigenvector. In this paper, we will first introduce some basic conceptes of matrix game and define a new concepts called optimal-dominated strategy, then we will introduce von Neumann's minimax theorem, which is the most important theorem in matrix game. With these propaedeutics, we will talk about the relationship between eigenvalue and properties of matrix game, we will first focus on three special cases in which the game matrices are diagonal, skew-symmetric and positive, then we will extend some conclusions to general matrices.

\section{Basic concepts of martix game}
\textbf{Definition 1.} Strategy set is the set of strategies player can choose when playing the game, denote strategy set of playerI and playerII as $S_1$ and $S_2$.
\\ \indent In matrix game, $|S_1|, |S_2|$ are finite.
\\ \indent  Every strategy in $S_1$ and $S_2$ is called pure strategy.

~\\
\textbf{Definition 2.} We call player I row player and player II column player, the game matrix $A \in R^{m \times n}$ is the matrix whose each row $i$ represent a strategy of row player and each column $j$ represent a strategy of column player, the entries $a_{ij}$ of game matrix represent the utility(earning) of row player when row player choose strategy $i$ and column player choose strategy $j$.
\\ \indent Matrix game is zero-sum game which means when strategies of row and column player are fixed, the sum of utility of these two players is zero. Thus $A$ corresponds to utility of row player and $-A$ corresponds to utility of column player.
~\\
\\Here is an example to illustrate this definition:
~\\Consider Rock-Paper-Scissors, here is its game matrix:
$$
\begin{pmatrix}
	0&-1&1\\
	1&0&-1\\
	-1&1&0
\end{pmatrix}
$$
\\ Rows represent rock, paper, scissors from top to bottom, columns represent rock, paper, scissors from left to right. When the row player wins, loses or the game draws, the corresponding entry of $A$ is $1,-1,0$.

~\\
\textbf{Definition 3.} Mixed strategy is a probablity distribution on strategy set, it can be represented by a stochastic vector. Let vector $x \in R^m$ be a mixed strategy of row player on $S_1$, and vector $y \in R^m$ be a mixed strategy of column player on $S_2$.
\\ \indent Mixed strategy vector describe the probability of choosing each pure strategy and combine these possibilities in a vector.
\\ \indent The utility of row player $u_1(x,y) = x^TAy$

~\\
\textbf{Definition 4.} Nash equilibrium: for a 2-player game, each play has a mixed strategy $\sigma_1, \sigma_2$, profile $(\sigma_1^*,\sigma_2^*)$ is called mixed strategy nash equilibrium if and only if $u_1(\sigma_1^*,\sigma_2^*) \ge u_1(\sigma_1, \sigma_2^*)$ and  $u_2(\sigma_1^*,\sigma_2^*) \ge u_2(\sigma_1^*, \sigma_2)$ for all $\sigma_1, \sigma_2$
\\ \indent Nash equilibrium is a kind of fixed point, at this point, no player can get more utility if it unilaterally deviates from original strategy. 
\\ \indent Nash equilibrium can be extended to n-player games

~\\
\textbf{Definition 5.} Maxmin value is $\max \limits_{x} \min \limits_{y}x^TAy$, it describes the largest value row player can get in the worst condition. Minmax value is $\min \limits_{y} \max \limits_{x}x^TAy$, it describes the largest value column player can get in the worst condition.

~\\
\textbf{Definition 6.} $x \in \arg\max \limits_{x} \min \limits_{y}x^TAy$ is called optimal strategy for row player. $y \in \arg\min \limits_{y} \max \limits_{x}x^TAy$ is called optimal strategy for column player. When a player choose the optimal strategy, its minimum utility is largest. 

~\\
\textbf{Definition 7.} We say a mixed strategy $x^*$ for row player is optimal-dominated if $\forall y, (x^*)^TAy = \max \limits_{x} \min \limits_{y}x^TAy$. Similarly, we say a strategy $y^*$ for column player is optimal-dominated if $\forall x, x^TAy^* = \min \limits_{y} \max \limits_{x}x^TAy$

~\\
\textbf{Lemma 1.} $x^*$ is optimal-dominated for row player if and only $(x^*)^TA = v$, where $v$ is a row vector whose entries are all $\max \limits_{x} \min \limits_{y}x^TAy$  

~\\
\emph{Proof:} Note that $y$ is stochastic vector, if $(x^*)^TA = v$, then $(x^*)^TAy = vy = \max \limits_{x} \min \limits_{y}x^TAy$. On the other hand, if $\forall y, (x^*)^TAy = \max \limits_{x} \min \limits_{y}x^TAy$, then let $y = e_1, e_2, \cdots, e_n$, we can deduce that each entry of $(x^*)^TAy$ is $\max \limits_{x} \min \limits_{y}x^TAy$, then $(x^*)^TA = v$.

~\\
Similarly, $y^*$ is optimal-dominated for column player if and only if $Ay = v.$ where $v$ is a column vector whose entries are all $\min \limits_{y} \max \limits_{x}x^TAy$ 

\section{von Neumann's minimax theorem}
Von Neumann's minimax theorem, from which we can infer that minmax value is equal to maxmin value, is the most important theorem in matrix game,

~\\
\textbf{Von Neumann's minimax theorem.} For any $A$, there is a mixed strategy $x^*$ for row player and $y^*$ for column player such that: $\quad \max\limits_{x}x^TAy^* = \min\limits_{y}(x^*)^TAy$
~\\
Besides, such $(x^*, y^*)$ is a nash equilibrium

~\\
This theorem has different proofs, Von Neumann first proved it in 1928$^{[1]}$ using Brouwer fixed point theorem, and then he published a more elegent proof using dual linear programming.$^{[2]}$

~\\
\textbf{Corollary 1. } $\max \limits_{x} \min \limits_{y}x^TAy = \min \limits_{y} \max \limits_{x}x^TAy = v(A)$

~\\
\emph{Proof:}
~\\
On the one hand: $\min \limits_{y} \max \limits_{x}x^TAy \le  \max\limits_{x}x^TAy^* =  \min\limits_{y}(x^*)^TAy \le \max \limits_{x} \min \limits_{y}x^TAy$
~\\
On the other hand, suppose $\min \limits_{y} \max \limits_{x}x^TAy =  \max \limits_{x}x^TAy'$. Note that $x^TAy' \ge \min\limits_{y}x^TAy$, then $\min \limits_{y} \max \limits_{x}x^TAy =  \max \limits_{x}x^TAy' \ge \max \limits_{x}\min\limits_{y}x^TAy$
~\\
Then we have $\max \limits_{x} \min \limits_{y}x^TAy = \min \limits_{y} \max \limits_{x}x^TAy$, define this value as $v(A)$.
\\ We also know that if $x^*, y^*$ satisfy $\max\limits_{x}x^TAy^* = \min\limits_{y}(x^*)^TAy$, then  $\max\limits_{x}x^TAy^* = \min\limits_{y}(x^*)^TAy = v(A)$. This will be used in the proof of Corollary 2.

~\\
\textbf{Corollary 2. } If $x', y'$ are optimal strategy, if and only if  $(x',y')$ is a nash equilibrium, and if $(x',y')$ is a nash equilibrium, $(x')^TAy'= v(A)$.

~\\
\emph{Proof:} If $x', y'$ are optimal strategies, then $v(A) = \min\limits_{y}(x')^TAy = \max\limits_{x}x^TAy'$, from Von Neumann's minimax theorem, $(x',y')$ is a nash equilibrium. 
\\ If $(x',y')$ is a nash equilibrium, then $u_1(x',y') = x'Ay' \ge xAy', \forall x,  u_2(x',y') = -u_1(x',y') = -x'Ay' \ge -x'Ay, \forall y.$ Then $\max \limits_{x} xAy' = x'Ay' = \min \limits_{y} x'Ay = v(A)$, thus $x'$ and $y'$ are optimal strategy. 

~\\
Another expression of corollary 2: \\ If $(x')^TAy'= v(A)$, then  $x', y'$ are optimal strategy. 
\\ This expression is useful in our discussion. 

\section{When $A$ is diagonal matrix}
\textbf{Lemma 2. }If $A$ has negative diagonal entries but not all diagonal values of $A$ is negative, then the optimal strategy vector $x'$ of row player has zero entries corresponding to negative diagonal values of $A$, and $v(A) = 0$.  

~\\
\emph{Proof:} When $x'$ has zero entries corresponding to negative diagonal values of $A$, $\min\limits_{y}(x')^TAy = 0$.  If $x'$ has any nonzero entry corresponding to negative diagonal value of $A$, Then $x'A$ must has negative entry because every entry of $x'$ is nonnegative, suppose this negative entry is in column $j$, then when column player choose pure strategy $j$, $u_1 < 0$, which means $\min\limits_{y}(x')^TAy < 0$, then such $x'$ can not be optimal strategy because $\min\limits_{y}(x')^TAy$ is maxmin value.

~\\
\textbf{Lemma 3. } If $A$ is not positive-definite or negative-definite , then $v(A) = 0$

~\\
\emph{Proof:} If $A$ has zero eigenvalue, which means at least one of digonal entries of $A$ is zero, then for any strategy $x$, $x^TA$ must have a zero entry. From lemma 2, $x^TA$ only has nonnegative entries, then $v(A) = 0$. If $A$ has negative eigenvalue but not negative-definite, then from lemma 2, $v(A) = 0$.

~\\
\textbf{Lemma 4. } If matrix $A$ is positive-definite and has positive eigenvalue $\lambda_1, \lambda_2, \cdots, \lambda_n$, then $v(A) = \frac{1}{\frac{1}{\lambda_1} + \frac{1}{\lambda_2} + \cdots + \frac{1}{\lambda_n}}$, and the optimal strategy of row player $x = (\frac{v(A)}{\lambda_1}, \frac{v(A)}{\lambda_2}, \cdots, \frac{v(A)}{\lambda_n})$. Besides, $x$ is optimal-dominated.

~\\
\emph{Proof:} Let $x=(x_1, x_2, \cdots, x_n)$, then $x^TA = (x_1\lambda_1, x_2\lambda_2, \cdots, x_n\lambda_n).$ If $x_1\lambda_1 = x_2\lambda_2 = \cdots = x_n\lambda_n$, combine these equations with $x_1+x_2+\cdots+x_n = 1$, we have $x_i = \frac{v}{\lambda_i}$, $v = \frac{1}{\frac{1}{\lambda_1} + \frac{1}{\lambda_2} + \cdots + \frac{1}{\lambda_n}}$. Now for any $y$,  $x^TAy = v$. If the equalities do not hold,  when $x_i$ increase, then there must exist $x_j$ whose value decrease. Denote the new strategy as $x'$, we have $\min\limits_{y}(x')^TAy \le x_j'\lambda_j < x_j\lambda_j = v$. When $x_i$ decrease,  Denote the new strategy as $x''$, we have $\min\limits_{y}(x'')^TAy \le x_i''\lambda_i < x_i\lambda_i = v$. Then $v(A) = \max\limits_{x} \min\limits_{y} x^TAy = v.$

~\\
Similarly, we can prove that when $A$ is negative-definite, the same conclusion holds. Also, for column player, we can get the same conclusion. 
Combine these three lemmas, we have a total understanding of the matrix game when $A$ is diagonal matrix. They are summaried in the following theorem.

~\\
\textbf{Theorem 1.} For a matrix game with diagonal game matrix $A$:
\\ For the value of $A$:
\\ (1) If $A$ is not positive-definite or negative definite, then $v(A) = 0$
\\ (2) If $A$ is positive-definite or negative-definite, then $v(A) = \frac{1}{\frac{1}{\lambda_1} + \frac{1}{\lambda_2} + \cdots + \frac{1}{\lambda_n}}$

~\\
For the optimal strategy:
\\ (1) If $A$ is not positive definite or negative definite, then the optimal strategy of row player has zero entries corresponding to negative diagonal values of $A$, the optimal strategy of column player has zero entries corresponding to negative diagonal values of $A$.
\\ (The proof of this conclusion is very similar to lemma 2)
~\\
\\ (2) If $A$ is positive definite or negative definite, then the optimal strategy of row and column player is $(\frac{v(A)}{\lambda_1}, \frac{v(A)}{\lambda_2}, \cdots, \frac{v(A)}{\lambda_n})$

\section{When A is skew-symmetric matrix}
It is common in real life that game matrix $A$ is skew-symmetric ($A^T = -A$), for example, Rock-Paper-Scissors.

~\\
\textbf{Theorem 2.} $v(A) = -v(-A^T)$.

~\\
\emph{Proof:} 

$
\begin{aligned}
v(A) &= \min \limits_{y} \max \limits_{x} x^TAy
\\ & = -(- \min \limits_{y} \max \limits_{x} x^TAy)
\\ & = -(\max \limits_{y} (-\max \limits_{x} x^TAy))
\\ & = -(\max \limits_{y} \min \limits_{x} (-x^TAy))
\\ & = -(\max \limits_{y} \min \limits_{x} y^T(-A^T)x)
\\ & = -v(-A^T)
\end{aligned}
$

~\\
~\\
\textbf{Corollary 3.} If game matrix $A$ is skew-symmetric, then $v(A) = 0$.

~\\
\emph{Proof: }
Using lemma 5, we know that $v(A) = -v(-A^T)$, here $A = -A^T$, then $v(A) = -v(A)$, $v(A) = 0$.

~\\
\textbf{Corollary 4.} If game matrix $A$ is skew-symmetric, then one player's optimal strategy is also optimal strategy for the other player.

~\\
\emph{Proof: }
In the proof of theorem 2, we have:  $\min \limits_{y} \max \limits_{x} x^TAy = -(\max \limits_{y} \min \limits_{x} y^T(-A^T)x)$, when $A$ is skew-symmetric,  $\min \limits_{y} \max \limits_{x} x^TAy = -(\max \limits_{y} \min \limits_{x} y^TAx) = 0$, then $\min \limits_{y} \max \limits_{x} x^TAy = \max \limits_{y} \min \limits_{x} y^TAx.$ Here optimal strategy of row player $y$ is also a optimal strategy of column player.

~\\
Now we have $v(A) = 0$ and from another expression of corollary 2, we know that if $(x')^TAy' = v(A)$, then  $x',y'$ are optimal strategy. To find optimal strategy of skew-symmetric matrix $A$, we need to find stochastic vector $x^*, y^*$ such that $(x^*)^TAy^* = 0$. It raises a problem that when $x^*$ or $y^*$ lies in the eigenspace of $A$. The following discussion will focus on this problem.

~\\
\textbf{Lemma 5.} The optimal strategy vector can only lies in the eigenspace of $A$ corresponding to zero eigenvalue.

~\\
\emph{Proof:} For optimal strategy $x^*, y^*$, $(x^*)^TAy = 0$. If $y$ is an eigen vector of $A$ corresponding to eigenvalue $\lambda$, then $\lambda x^Ty = 0$. If $\lambda \neq 0$, then $x^Ty = 0$, but this is impossible because $x^*$ and $y^*$ are stochastic vectors. If $x$ is an eigenvector of $A$ corresponding to eigenvalue $\lambda$, then $Ax = \lambda x, x^TA^T = \lambda x^T.$ Then $x^TAy = -x^TA^Ty = -\lambda x^Ty = 0$. Similarly, $x^Ty$ can not be zero, then $\lambda = 0$.

~\\
This lemma shows that if optimal strategies are eigenvectors of $A$, then thay must lie in the null space of $A$. Now we need to investigate the properties of null space of skew-symmetric matrix $A$.
What makes a difference here is that $x$ is a stochastic vector, so our problem now is when do $A$ has a stochastic vector in its null space.
Here is a general theorem related to this problem.

~\\
\textbf{Gordan's theorem:} Given a matrix $A$, the following are alternatives:
\\ (1) $Ax = 0, x \ge 0$ has a solution $x$
\\ (2) $A^Ty > 0$ has a solution $y$.
\\ This theorem is proved by Gordan in 1873.$^{[3]}$

~\\
\textbf{Theorem 3.} If $A$ is a skew-symmetric matrix, then there are optimal strategies in eigenspace of $A$ if and only if $Ay > 0$ has a solution $y$.

~\\
\emph{Proof:} By lemma 5, we know that if optimal strategy $z^*$ in eigenspace of $A$, then $Az^* =0$. By Gordan's theorem, this equation has nonnegative solution $z^*$ if and only if $A^Ty > 0$ has a solution $y$. Here $A$ is skew-symmetric, then the condition is equial to $Ay > 0$ has a solution $y$. 
\\
 \indent  With nonegative solution $z^*$, we can construct stochastic vector easily by letting the sum of all entries devide each entry.

\section{When A is positive matrix}
When every entry of a matrix/vector is positive, then it is positive matrix/vetcor. There is a remarkable theorem for such matrix which is useful for our analysis.

~\\
\textbf{Perron–Frobenius theorem:}  If $A$ is a positive matrix, then it has the following properties:
\\ (1) It has a positive real eigenvalue $\lambda^*$ which has the largest module among all eigenvalues of $A$. 
\\ (2) There is an eigenvector $x \in E_{A,\lambda^*}$ such that every entry of $x$ is positive. 
\\ (3) In $E_{A,\lambda}$, there is no other positive vector except positive times of $x$.
\\ The proof of these conclusions can be found in Meyer.$^{[4]}$

~\\
\textbf{Lemma 6.} If there is an optimal strategy vector $y'$, whose entries are all positive, for column player. Then any optimal strategy of row player is optimal-dominated.

~\\
\emph{Proof:} Suppose the game matrix is $A$, for any row-player optimal strategy $(x^*)^T$, we have $(x^*)^TAy' = v(A)$. If $(x^*)^TA$ has an entry larger than $v(A)$ then the corresponding entry of $y'$ will be zero. If $(x^*)^TA$ has an entry smaller than $v(A)$, as $y'$ minimize $(x^*)^TA$, $(x^*)^TAy'$ will be smaller than $v(A)$.\\ In that case, if there is an positive optimal strategy vector $y'$, then for any optimal strategy $x^*$, $(x^*)^TA$ must equal to the vector whose all entries are $v(A)$, which means any optimal strategy of row player is optimal-dominated.

~\\
Similarly, we can get such conclusion: 
\\ If there is an optimal strategy vector $x'$, whose entries are all positive, for row player. Then any optimal strategy of column player is optimal-dominated.

~\\
\textbf{Theorem 4:} For positive matrix $A \in R^{n \times n}$ with the largest eigenvalue $\lambda^*$ and unique positive stochastic vector $y^*$ in $E_{A,\lambda^*}$, if $ \lambda^* min_{i \in [n]} \{y_i^*\}\le v(A) \le \lambda^* max_{i \in [n]} \{y_i^*\}$, then any optimal strategy of row player is optimal-dominated.

~\\
\emph{Proof: } From Perron–Frobenius theorem, we know that there exists an unique positive stochastic vector $y^*$ in $E_{A, \lambda^*}$. For any optimal strategy $x^*$ of row player, $v(A) = (x^*)^TAy^* = (x^*)^T \lambda^* y^* $.\\  Note that $(x^*)^Ty^* = x_1^*y_1^* + \cdots + x_n^*y_n^*$ and $\sum_{i \in [n]} x_i^* = 1$,  $(x^*)^Ty^*$ is a convex combination of $y_i^*$, then $ min_{i \in [n]} \{y_i^*\} \le (x^*)^Ty^* \le max_{i \in [n]} \{y_i^*\}$. 
\\For any value $v$ in $[min_{i \in [n]} \{y_i^*\}, max_{i \in [n]} \{y_i^*\}]$, there is an $x'$ such that $(x')^Ty^* = v$. Then if $ \lambda^* min_{i \in [n]} \{y_i^*\}\le v(A) \le \lambda^* max_{i \in [n]} \{y_i^*\}$, there is an $x'$ such that $(x')^TAy^* = v(A)$. Then $y^*$ is an optimal strategy, then by lemma 6, any optimal strategy of row player is optimal-dominated.

~\\  
Note that if $A$ is positive with largest eigenvalue $\lambda^*$, then $A^T$ is positive with largest eigenvalue $\lambda^*$, thus there is an unique stochastic vector $x^*$ such that $(x^*)^TA = \lambda^* (x^*)^T$
~\\
\\ Similarly, we can get such conclusion:
\\ For positive matrix $A \in R^{n \times n}$ with the largest eigenvalue $\lambda^*$ and unique positive stochastic vector $x^*$ in $E_{A^T,\lambda^*}$, if $ \lambda^* min_{i \in [n]} \{x_i^*\}\le v(A) \le \lambda^* max_{i \in [n]} \{x_i^*\}$, then any optimal strategy of column player is optimal-dominated.

\section{When A is general matrix}
Some conclusions we discussed above can be extended to general case with some modification. 
\\ Note that when we discuss skew-symmertic martix $A$, if $\lambda = 0$ is an eigenvalue of $A$, then $x^TAy = 0 = v(A)$, where $x,y$ are optimal strategies and also in $E_{A,0}$
\\ We extend this conclusion to a more general form:

~\\
\textbf{Theorem 4.} For any square matrix $A$ with real eigenvalue $\lambda$, if there are stochastic vectors $x^*,y^*$ such that $y^* \in E_{A,\lambda}, x^* \in E_{A^T,\lambda}$, then $v(A-\lambda I) = 0.$ And $x^*,y^*$ are optimal strategy (also optimal-dominated)

~\\
\emph{Proof:} On the one hand, $A^Tx^* = \lambda x^* \Rightarrow (x^*)^TA = \lambda (x^*)^T \Rightarrow (x^*)^T(A-\lambda I) = 0 \Rightarrow \min\limits_{y}(x^*)^TAy \Rightarrow \max \limits_{x}\min\limits_{y}x^TAy \ge 0.$ 
\\
\indent On the other hand, $Ay^* = \lambda y^* \Rightarrow (A-\lambda)y^* = 0 \Rightarrow \max\limits_{x}x^TAy^* = 0 \Rightarrow \min \limits_{y}\max\limits_{x}x^TAy \le 0.$ Then $v(A-\lambda I) = 0$

\section{Conclusion}
Although matrix game module has come out for a long time, it still has many interesting properties to be investigated. In this paper, we mainly focus on three special cases to examine the relationship between eigenvalue/eigenvector and matrix game, but there should be more relation between them. In future research, skew-matrix is also a good materials, note that an $n \times n$ matrix with odd $n$ always has zero eigenvalue, this may be useful to investigate its null space. Also, Courant-Fischer theorem has similar maxmin form with our discussion's, if this theorem can be applied in some way, there will be far more relation of eigenvalue/eigenvector and matrix game.


%

\ifCLASSOPTIONcaptionsoff
  \newpage
\fi

\end{document}